\PassOptionsToPackage{table}{xcolor}

\documentclass[sigconf,hyphens]{acmart}

\usepackage{glossaries}
\usepackage{cleveref}
\usepackage{multirow}        
\usepackage{url}     
\usepackage{breakurl}         
\usepackage{hyperref}         
\usepackage[T1]{fontenc}      
\usepackage{microtype}        

\hypersetup{breaklinks=true}

\usepackage{subfigure}

\AtBeginDocument{%
  }

\copyrightyear{2025}
\acmYear{2025}
\setcopyright{rightsretained}
\acmConference[CF Companion '25]{22nd ACM International Conference on
Computing Frontiers}{May 28--30, 2025}{Cagliari, Italy}
\acmBooktitle{22nd ACM International Conference on Computing Frontiers (CF Companion '25), May 28--30, 2025, Cagliari,
Italy}\acmDOI{10.1145/3706594.3726974}
\acmISBN{979-8-4007-1393-4/2025/05}

\newcommand\eg{e.g.,\ }
\newcommand\ie{i.e.,\ }

\newacronym[longplural={Scratchpad Memories}]{SPM}{SPM}{Scratchpad Memory}
\newacronym{ACE}{ACE}{AXI Coherent Extensions}
\newacronym{ADDRGEN}{ADDRGEN}{address generator}
\newacronym{AMBA}{AMBA}{Advanced Microcontroller Bus Architecture}
\newacronym{APB}{APB}{Advanced Peripheral Bus}
\newacronym{API}{API}{Application Programming Interface}
\newacronym{ASIC}{ASIC}{Application-Specific Integrated Circuit}
\newacronym{AVX}{AVX}{Advanced Vector Extension}
\newacronym{AXI}{AXI}{Advanced eXtensible Interface}
\newacronym{BLAS}{BLAS}{Basic Linear Algebra Subprograms}
\newacronym{CHI}{CHI}{Coherent Hub Interface}
\newacronym{CMOS}{CMOS}{Complementary Metal-Oxide-Semiconductor}
\newacronym{CNN}{CNN}{Convolutional Neural Network}
\newacronym{CPU}{CPU}{Central Processing Unit}
\newacronym{CSR}{CSR}{control and status register}
\newacronym{CTS}{CTS}{Clock Tree Synthesis}
\newacronym{DLP}{DLP}{Data Level Parallelism}
\newacronym{DMA}{DMA}{Direct Memory Access}
\newacronym{DRAM}{DRAM}{Dynamic Random-Access Memory}
\newacronym{DSP}{DSP}{Digital Signal Processing}
\newacronym{DUT}{DUT}{Device Under Test}
\newacronym{EW}{EW}{element width}
\newacronym{EEW}{EEW}{Effective Element Width}
\newacronym{ECL}{ECL}{Emitter-Coupled Logic}
\newacronym{FBB}{FBB}{Forward Body-Biasing}
\newacronym{FDSOI}{FD-SOI}{Fully Depleted Silicon-on-Insulator}
\newacronym{FMA}{FMA}{Fused Multiply-Add}
\newacronym{FSM}{FSM}{finite state machine}
\newacronym{FPGA}{FPGA}{Field-Programmable Gate Array}
\newacronym{FP}{FP}{Floating Point}
\newacronym{FPU}{FPU}{Floating Point Unit}
\newacronym{GPGPU}{GPGPU}{General-Purpose \acrlong{GPU}}
\newacronym{GPU}{GPU}{Graphics Processing Unit}
\newacronym{HDL}{HDL}{Hardware Description Language}
\newacronym{HERO}{HERO}{Heterogeneous Embedded Research Platform}
\newacronym{HPC}{HPC}{high-performance computing}
\newacronym{ILP}{ILP}{Instruction Level Parallelism}
\newacronym{IoT}{IoT}{Internet-of-Things}
\newacronym{IOT}{IoT}{Internet-of-Things}
\newacronym{IPC}{IPC}{Instructions Per Cycle}
\newacronym{IPU}{IPU}{Image Processing Unit}
\newacronym{SoC}{SoC}{system on chip}
\newacronym{ISA}{ISA}{instruction set architecture}
\newacronym{LSB}{LSB}{Least Significant Bit}
\newacronym{LSU}{LSU}{Load/Store Unit}
\newacronym{LVT}{LVT}{low voltage threshold}
\newacronym{MIMD}{MIMD}{multiple instruction, multiple data}
\newacronym{MMU}{MMU}{memory management unit}
\newacronym{MUL}{MUL}{multiplier}
\newacronym{ML}{ML}{machine learning}
\newacronym{NoC}{NoC}{network on chip}
\newacronym{MVL}{MVL}{maximum vector length}
\newacronym{NUMA}{NUMA}{non-uniform memory access}
\newacronym{NOC}{NoC}{Network-on-Chip}
\newacronym{PCIe}{PCIe}{Peripheral Component Interconnect Express}
\newacronym{PC}{PC}{Program Counter}
\newacronym{PE}{PE}{processing element}
\newacronym{PL}{PL}{Programmable Logic}
\newacronym{PMCA}{PMCA}{Programmable Manycore Accelerator}
\newacronym{PPA}{PPA}{power, performance, area}
\newacronym{NZE}{NZE}{non-zero element}
\newacronym{NER}{NER}{non-empty row}
\newacronym{LLC}{LLC}{last-level cache}
\newacronym{PSL}{PSL}{Power Service Layer}
\newacronym{LRU}{LRU}{least-recently-used}
\newacronym[longplural=page table entries]{PTE}{PTE}{page-table entry}
\newacronym{PTW}{PTW}{page-table walker}
\newacronym{PULP}{PULP}{Parallel Ultra Low Power}
\newacronym{RAW}{RAW}{read-after-write}
\newacronym{RBB}{RBB}{Reverse Body-Biasing}
\newacronym{ROB}{ROB}{Reorder Buffer}
\newacronym{TLB}{TLB}{translation lookaside buffer}
\newacronym{DTLB}{DTLB}{data translation lookaside buffer}
\newacronym{RTL}{RTL}{Register Transfer Level}
\newacronym{VM}{VM}{virtual memory}
\newacronym{RVT}{RVT}{Regular Voltage Threshold}
\newacronym{RoCC}{RoCC}{Rocket Custom Coprocessor Interface}
\newacronym{SCM}{SCM}{Storage Class Memory}
\newacronym{SEW}{SEW}{Single Element Width}
\newacronym{SIMD}{SIMD}{single instruction, multiple data}
\newacronym{SIMT}{SIMT}{single instruction, multiple thread}
\newacronym{SLDU}{SLDU}{Slide Unit}
\newacronym{SLVT}{SLVT}{super-low voltage threshold}
\newacronym{SM}{SM}{Streaming Multiprocessor}
\newacronym[longplural={Static Random-Access Memories}]{SRAM}{SRAM}{Static Random-Access Memory}
\newacronym{SSE}{SSE}{Streaming SIMD Extension}
\newacronym{SVE}{SVE}{Scalable Vector Extension}
\newacronym{TLP}{TLP}{Thread Level Parallelism}
\newacronym{TxnID}{TxnID}{Transaction ID}
\newacronym{VAC}{VAC}{Vector Access}
\newacronym{VC}{VC}{virtual channel}
\newacronym{VCONV}{VCONV}{Vector Conversion}
\newacronym{VEX}{VEX}{Vector Execute}
\newacronym{VFU}{VFU}{vector functional unit}
\newacronym{VID}{VID}{Vector Instruction Decode}
\newacronym{VIS}{VISSUE}{Vector Instruction Issue}
\newacronym{VLIW}{VLIW}{Very Long Instruction Word}
\newacronym{VLOOP}{VLOOP}{Vector Loop}
\newacronym{VLR}{VLR}{vector length register}
\newacronym{VLSU}{VLSU}{vector load-store unit}
\newacronym{VNB}{VNB}{Von Neumann Bottleneck}
\newacronym{VRF}{VRF}{vector register file}
\newacronym{VT}{VT}{vector thread}
\newacronym{BW}{BW}{bandwidth}
\newacronym{MASKU}{MASKU}{Mask Unit}
\newacronym{VU0.5}{VU0.5}{Vector Unit 0.5}
\newacronym{VU1.0}{VU1.0}{Vector Unit 1.0}
\newacronym{VMFPU}{VMFPU}{Vector Multiplier/Floating Point Unit}
\newacronym{VFPU}{VFPU}{Vector Floating Point Unit}
\newacronym{VDIV}{VDIV}{Vector Divider}
\newacronym{VMUL}{VMUL}{Vector Multiplier}
\newacronym{WAR}{WAR}{write-after-read}
\newacronym{WAW}{WAW}{write-after-write}
\newacronym{DCT}{DCT}{discrete cosine transform}
\newacronym{TSV}{TSV}{through-silicon via}
\newacronym{3DIC}{3D-IC}{three-dimensional integrated circuit}
\newacronym{F2F}{F2F}{face-to-face}
\newacronym{IC}{IC}{integrated circuit}
\newacronym{C4}{C4}{controlled collapse chip connection}
\newacronym{FEOL}{FEOL}{front end of the line}
\newacronym{OS}{OS}{operating system}
\newacronym{BEOL}{BEOL}{back end of the line}
\newacronym{SLEN}{SLEN}{striping distance}
\newacronym{VSU}{VSU}{Vector Store Unit}
\newacronym{DNN}{DNN}{Deep Neural Networks}
\newacronym{AI}{AI}{Artificial Intelligence}
\newacronym{AR}{AR}{Augmented Reality}
\newacronym{SoA}{SoA}{State-of-the-Art}
\newacronym{FD-SOI}{FD-SOI}{Fully Depleted - Silicon on Insulator}
\newacronym{RVV}{RVV}{RISC-V ``V''}
\newacronym{ALU}{ALU}{Arithmetic-Logic Unit}
\newacronym{FFT}{FFT}{Fast Fourier Transform}
\newacronym{VLDU}{VLDU}{Vector Load Unit}

\begin{document}

\title{AraOS: Analyzing the Impact of Virtual Memory Management on Vector Unit Performance}

\author{Matteo Perotti}
\email{mperotti@iis.ee.ethz.ch}
\orcid{}
\affiliation{%
  \institution{ETH Zürich}
  \city{Zürich}
  \state{}
  \country{Switzerland}
}

\author{Vincenzo Maisto}
\email{vincenzo.maisto2@unina.it}
\affiliation{%
  \institution{Università di Napoli Federico II}
  \city{Napoli}
  \country{Italy}
}

\author{Moritz Imfeld}
\email{moimfeld@student.ethz.ch}
\affiliation{%
  \institution{ETH Zürich}
  \city{Zürich}
  \country{Switzerland}
}

\author{Nils Wistoff}
\email{nwistoff@iis.ee.ethz.ch}
\affiliation{%
  \institution{ETH Zürich}
  \city{Zürich}
 \state{}
 \country{Switzerland}}
 
\author{Alessandro Cilardo}
\email{acilardo@unina.it}
\affiliation{%
 \institution{Università di Napoli Federico II}
 \city{Napoli}
 \state{}
 \country{Italy}
}

\author{Luca Benini}
\email{lbenini@iis.ee.ethz.ch}
\affiliation{%
  \institution{ETH Zürich - Università di Bologna}
  \city{Zürich, Switzerland - Bologna, Italy}
  \state{}
  \country{}}
  
\renewcommand{\shortauthors}{M. Perotti, V. Maisto, M. Imfeld, N. Wistoff et al.}

\begin{abstract}
Vector processor architectures offer an efficient solution for accelerating data-parallel workloads (\eg ML, AI), reducing instruction count, and enhancing processing efficiency. This is evidenced by the increasing adoption of vector ISAs, such as Arm’s SVE/SVE2 and RISC-V’s RVV, not only in high-performance computers but also in embedded systems.
The open-source nature of RVV has particularly encouraged the development of numerous vector processor designs across industry and academia.
However, despite the growing number of open-source RVV processors, there is a lack of published data on their performance in a complex application environment hosted by a full-fledged operating system (Linux).
In this work, we add OS support to the open-source bare-metal Ara2 vector processor (AraOS) by sharing the MMU of CVA6, the scalar core used for instruction dispatch to Ara2, and integrate AraOS into the open-source Cheshire SoC platform. 
We evaluate the performance overhead of virtual-to-physical address translation by benchmarking matrix multiplication kernels across several problem sizes and \gls{TLB} configurations in CVA6's shared MMU, providing insights into vector performance in a full-system environment with virtual memory. With at least 16 TLB entries, the virtual memory overhead remains below 3.5\%.
Finally, we benchmark a 2-lane AraOS instance with the open-source RiVEC benchmark suite for RVV architectures, with peak average speedups of 3.2$\times$ against scalar-only execution. 
\end{abstract}



\maketitle

\section{Introduction}

The diffusion of \gls{ML} applications, with algorithms evolving at a rapid pace, necessitates hardware architectures that can quickly adapt via highly productive software programming. 
Furthermore, energy efficiency is key in prolonging battery life, reducing cooling demands, and minimizing electricity consumption.

Scalable vector processors offer an efficient architectural solution for accelerating data-parallel computation and, in particular, \gls{ML} workloads while ensuring flexibility and a well-understood, streamlined programming model. 
These architectures leverage \gls{SIMD} computation to reduce instruction fetch overhead, process multiple elements in parallel, and improve data reuse closer to the processing units \cite{ara2perotti24}.

The growing interest in scalable vector architectures is evident in industry and academia and across numerous application domains.
For instance, Arm's SVE \gls{ISA} powers the FUGAKU supercomputer, which dominated the TOP500 supercomputer performance ranking for two years until mid-2022. The new version of SVE, SVE2, targets general-purpose software beyond \gls{HPC} and \gls{ML} and is supported by the Arm Neoverse V2, featured in the NVIDIA Grace CPU \cite{nvidiagrace}.
Furthermore, the open-source RISC-V \gls{ISA} ratified its \gls{RVV} extension in 2021, and multiple companies are already supporting it with their processors \cite{SiFiveX280, Xuantie2020, Andesax45mpv, Semidynamics}. Notably, the majority of commercial processors support an \gls{OS}, leveraging vectors not only for numerical code but also to accelerate general-purpose functions such as \texttt{memcpy} and string operations.  

Since its first drafts, \gls{RVV} has also been adopted in multiple open-source \cite{zhao2024instructionschedulingsaturnvector, ara2perotti24, platzer2021, risc-v-squared} and closed-source non-commercial designs \cite{Minervini2022, maisto2022}, and several of them demonstrated competitive power, performance, and area metrics. However, to the best of our knowledge, none of the open-source \gls{RVV} architectures has reported \gls{OS} support or performance evaluation in an \gls{OS} environment with full support for virtual memory, which is essential for the majority of applications.

In this paper, we present AraOS, the first open-source \gls{RVV} vector processor architecture with full \gls{OS} support, along with a performance evaluation when executing key vector kernels under Linux with virtual memory. We deploy AraOS on a \gls{FPGA} with an open-source flow, enabling rapid prototyping and further research into \gls{OS}-integrated vector processing.

In a nutshell, the contributions of this work are:
\begin{itemize}
\item We present AraOS, to the best of our knowledge, the first open-source\footnote{The code of AraOS is available at https://github.com/pulp-platform/ara} RISC-V V 1.0 vector processor supporting an \gls{OS}. AraOS builds on the bare-metal Ara2's architecture \cite{ara2perotti24} and features dedicated logic and an additional interface to support virtual memory, incurring only 2.4\% area overhead.
\item We integrate AraOS in the Cheshire open-source platform to study the impact of the \gls{OS} and address translation on the system's performance depending on the application vector length and \gls{TLB} size. 
With at least 16 TLB entries, the virtual memory performance overhead does not exceed 3.5\%. 
\item We synthesize the architecture on a Xilinx VCU128 \gls{FPGA} using an open-source flow and characterize its performance using the open-source RiVEC benchmark suite \cite{10.1145/3422667}. Cheshire powered by a 2-lane AraOS is 3.2$\times$ faster and +39\% more area efficient than its scalar baseline on average. 
\end{itemize}

AraOS aims to become the reference design for application processors with \gls{RVV}-1.0 vector support, a baseline for benchmarking and optimizing future open- and closed-source implementations.

\section{Architecture}
\begin{figure}[ht]
    \centering
    \includegraphics[width=1\linewidth]{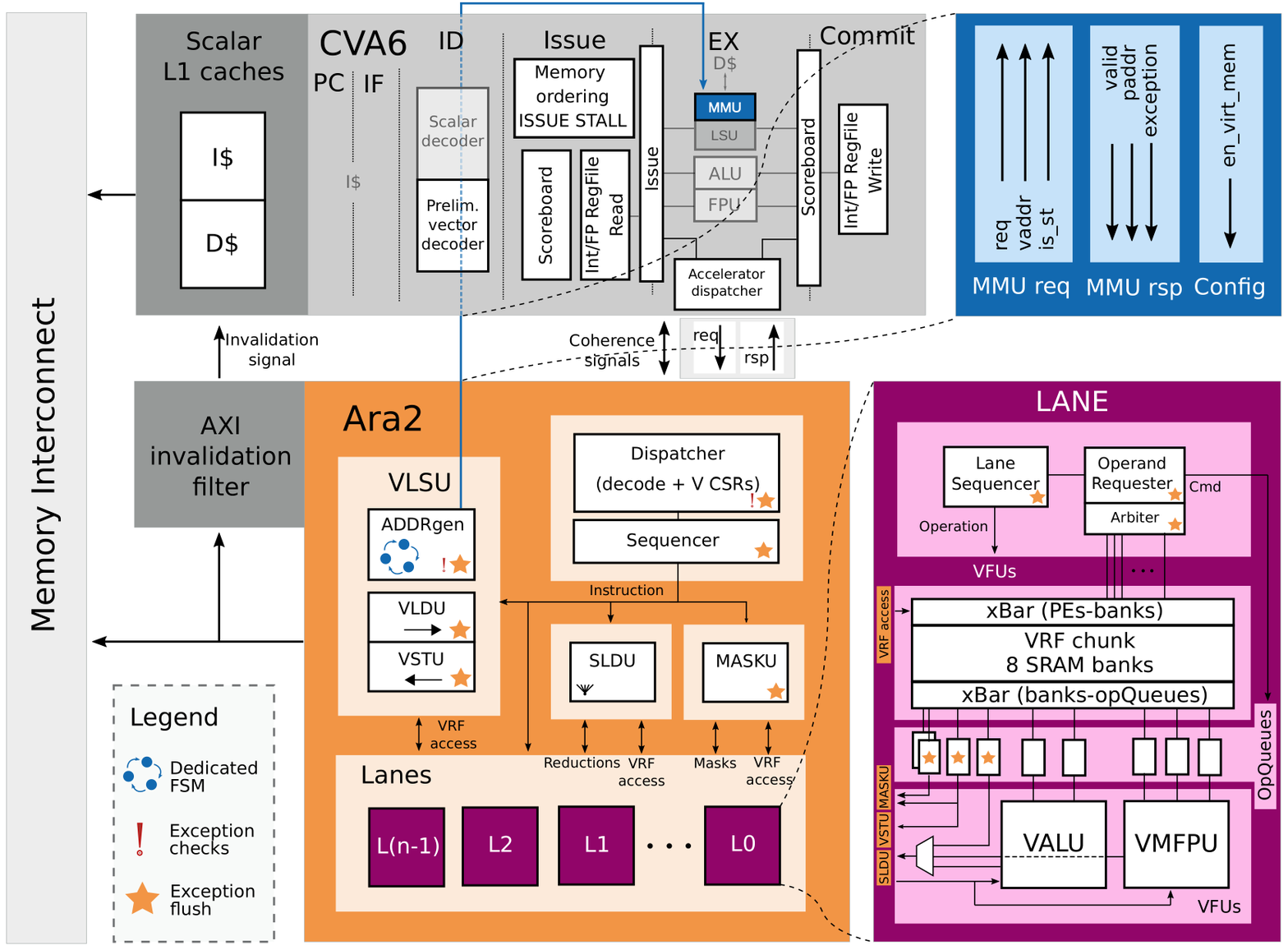}
    \caption{AraOS with shared MMU to enable virtual memory.}
    \label{fig:arch_diag}
    \Description[Schematic of AraOS.]{Schematic of AraOS.}
\end{figure}

AraOS is a decoupled architecture based on the Ara2 RVV-1.0 vector accelerator and the scalar application-class core CVA6 (\texttt{RV64GC}). AraOS features additional logic to support virtual memory alongside CVA6, which natively supports it.
To enable virtual memory support in Ara2's private \gls{VLSU}, we add a dedicated \gls{MMU} interface between Ara2's \gls{ADDRGEN} and CVA6's MMU. The vector unit translates virtual addresses to physical ones by using CVA6's MMU, which is time-shared between the core and the accelerator.
\Cref{fig:arch_diag} shows the architecture of AraOS.

\textbf{ADDRGEN.}
Ara2's \gls{VLSU} uses the AXI protocol for memory requests. Before issuing requests to memory and if virtual memory is enabled, the \gls{ADDRGEN} requests a virtual-to-physical address translation to CVA6's \gls{MMU} for each AXI \texttt{AR}/\texttt{AW} transaction. 
Ara2 optimizes unit-strided vector memory operations through AXI bursts limited by 4-KiB page boundaries (as required by the AXI specifications \cite{axi}), minimizing the number of \gls{MMU} requests with only one translation per burst.
When CVA6's \gls{MMU} receives a request, dedicated multiplexing logic connects the \gls{MMU} to the requester until the \gls{MMU} responds.

\textbf{MMU interface.}
The \gls{MMU} interface, shown in \Cref{fig:arch_diag}, includes a \texttt{req} signal to start the request, with the target virtual address \texttt{vaddr} and related metadata for \gls{MMU} exception checks.
The \gls{MMU} responds with a \texttt{valid} signal to indicate that the physical address \texttt{paddr} and the information about possible exceptions are valid.
CVA6 enables Ara2's virtual memory through the \texttt{satp} CSR.

\textbf{Exception handling.}
CVA6 dispatches every vector instruction to Ara2 non-speculatively when it reaches the top of the scoreboard and, before committing, waits for an answer that arrives a) as soon as Ara2 has checked that no exception occurred and the result (if needed) is valid or b) if an exception occurs. 
All the exceptions related to vector instructions are handled by CVA6 at commit time. 
With the addition of virtual memory, accessing memory pages can generate page-fault exceptions in the middle of vector memory instructions. 
When this happens, the \gls{ADDRGEN} does not issue any further translation requests to the \gls{MMU}. Ara2 reports the exception to CVA6 and starts its post-exception flushing procedure.

The exception notice is propagated through Ara2's frontend, which is flushed, and the index of the faulty element is saved into the \texttt{vstart} CSR.
However, the backend can still be in the process of committing valid results from previous instructions or vector elements before the faulty one from the same instruction. At the same time, the faulty instruction has dirtied the micro-architectural state, \eg by fetching source operands from the \gls{VRF} to the operand queues.
To successfully orchestrate the flushing procedure, a dedicated \gls{FSM} stalls the frontend until all the operations preceding the one on the faulty element have committed their architectural state.
Then, the \gls{FSM} injects into the backend a flush request that propagates through the pipeline flushing the micro-architectural state (registers, FIFOs, \glspl{FSM}) until it acknowledges back the frontend \gls{FSM} after $\sim$10 cycles, allowing normal operation again. This procedure is not latency-critical, as it only happens when dealing with a page-fault exception.

\textbf{SoC integration.}
We integrate AraOS in the Cheshire open-source \gls{SoC} \cite{ottaviano2023cheshire}, as shown in \Cref{fig:che_diag}. 
Cheshire is a modular system that already supports CVA6 as a host core. Its main AXI crossbar handles 64-bit wide data buses and connects the core with the \gls{LLC} (then connected to an external DRAM), a boot ROM, and peripherals, including UART and JTAG. 
Ara2's parametric AXI memory port passes through a downsizer to adapt its data width to 64 bits before entering the crossbar. 

\textbf{Cache coherence.}
To ensure coherence, CVA6 uses a write-through L1 data cache, so that Ara2 can always access up-to-date data from the \gls{LLC}. 
Also, a dedicated AXI invalidation filter monitors the physical addresses on Ara2's \texttt{AW} AXI channel. Upon a vector store, the filter sequentially flushes CVA6's L1 cache sets addressed by the physical indexes accessed by the vector store AXI burst. Despite CVA6's L1 data cache being virtually indexed, this mechanism remains effective when virtual memory is enabled as long as the number of L1 cache sets is at most 4096 since the least significant 12 bits of every address are also physical. CVA6's data cache has 2048 sets (8-KiB, 4-way associative) and only requires an 11-bit index, which is, therefore, always physical. 
Finally, CVA6 and Ara2 are mutually aware of the status of their in-flight memory operations to enforce coherence and avoid violating memory ordering \cite{ara2perotti24}.

\section{Experiments}
To evaluate the impact of the added hardware support for virtual memory (\gls{MMU} interface and multiplexing, \gls{VLSU} additional logic, and post-page-fault flush FSM) on CVA6 and Ara2, we synthesize and place-and-route AraOS configured with two lanes in 22-nm technology as in \cite{ara2perotti24}. The additional hardware does not impact the worst-case maximum frequency of the architecture (950 MHz, 0.72V, 125C, SS), and the area overhead is limited to 45 kGE (+2.4\%).

Then, we integrate AraOS in the Linux-ready Cheshire \gls{SoC} platform~\cite{ottaviano2023cheshire} with 1-MiB \gls{LLC} and no VGA, and deploy it on an AMD Xilinx VCU128 board~\cite{vcu128} running at 50 MHz. 
AraOS is configured with two lanes, \ie with two \glspl{FPU}, VLEN$=$2048, and a memory bandwidth of 64 bits/cycle.
Adding Ara2 (1.33 M LUTs) to the Cheshire \gls{SoC} (1.02 M LUTs) incurs a $2.3\times$ footprint.
To measure the impact of interrupts and context switches, we add a measurement infrastructure composed of performance counters and FIFOs to create snapshots of the internal state of the architecture and relevant event timestamps.
In the following, we characterize AraOS's performance on FPGA while running Linux kernel 6.5 and compile all vector benchmarks with \texttt{gcc} 13.2.0 (\texttt{-O3}).

\subsection{OS Overhead Evaluation}
\label{subsection:soc:araos:evaluation}
To evaluate the \gls{OS} overhead on vector applications, we measure the performance of matrix multiplication benchmarks running as Linux applications against their bare-metal baselines while varying CVA6's \gls{DTLB} size from 2 to 128 \glspl{PTE}. We choose matrix multiplication as an example of a vector kernel that heavily requires the cooperation of the scalar core, interleaving scalar and vector memory requests.

\textbf{\gls{OS} scheduler.}
We assess the impact of the \gls{OS} scheduler's interrupts and context switches with default settings. Interrupts that wake up the scheduler are fired at 100Hz and take $\sim$20k cycles to get back to the vector process when there is no context switch. During a context switch between two vector processes, the vector state (\gls{VRF} and vector CSRs) is saved to memory and then restored. This takes $\sim$3.2k cycles, which is in line with the expectations, as a context switch between two scalar processes takes $\sim$1k cycles, and AraOS needs $\sim$2k cycles to save and restore its 8-KiB \gls{VRF} with a 64-bit/cycle memory BW.
The \gls{TLB} and cache pollution overhead due to the scheduler intervention is always below 0.5\% of the runtime. 

\textbf{Virtual memory.}
We assess the performance impact of address translation on matrix multiplication kernels against various problem and \gls{DTLB} sizes with a non-preemptive scheduler, \ie the running process cannot be interrupted. 
\begin{figure*}[ht]
    \centering
    \subfigure[AraOS integration in the Cheshire SoC.] {
        \includegraphics[width=0.3\linewidth,trim=4 4 4 2,clip]{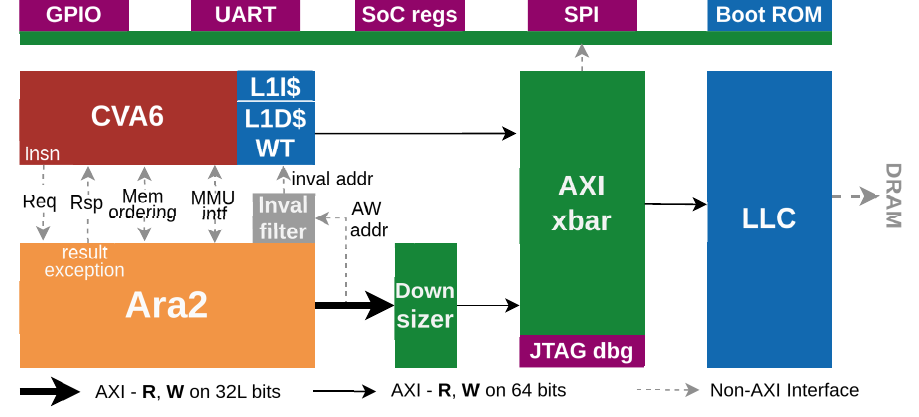}
        \label{fig:che_diag}
    }
    \subfigure[Matmul 32x32] {
        \includegraphics[width=0.2\linewidth]{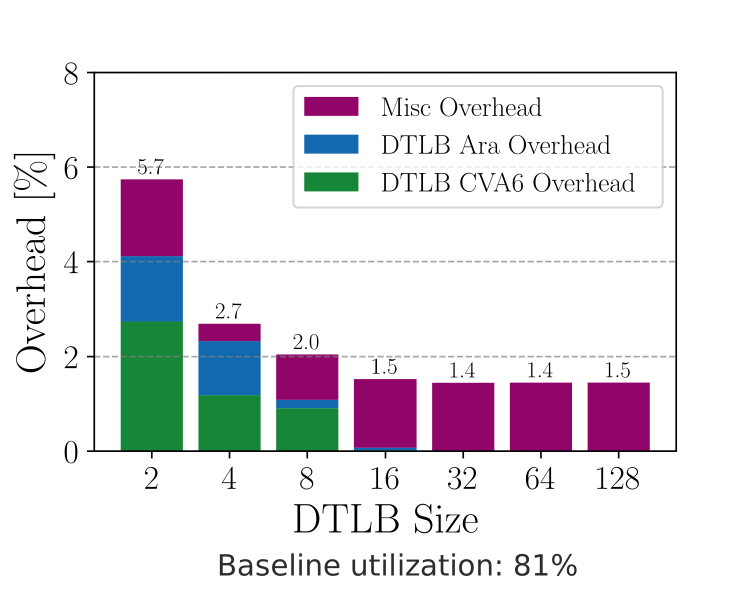}
    }
    \subfigure[Matmul 64x64] {
        \includegraphics[width=0.2\linewidth]{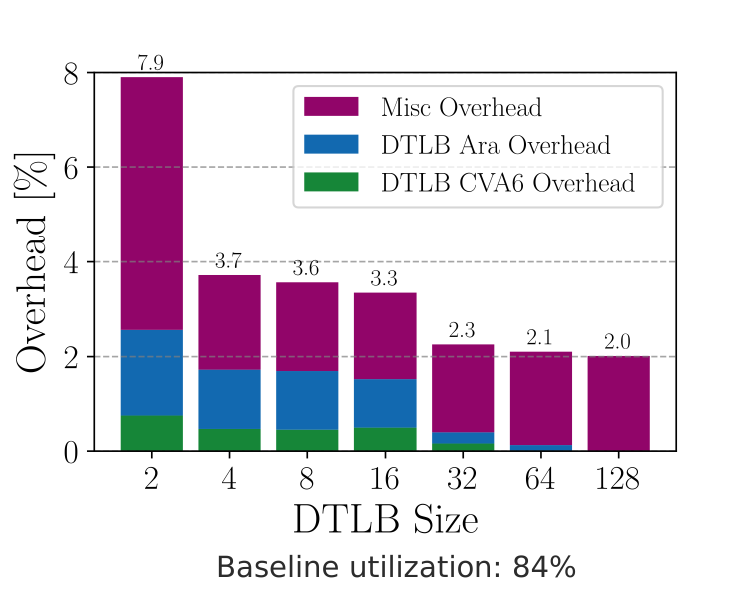}
    }
    \subfigure[Matmul 128x128] {
        \includegraphics[width=0.2\linewidth]{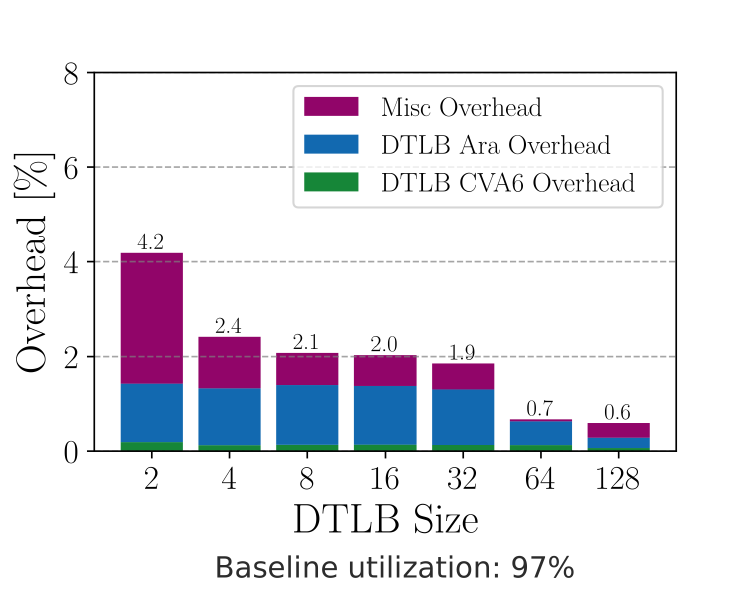}
    }
    \caption{AraOS integration in Cheshire SoC and \gls{OS} perf. overhead for a matrix multiplication on different problem sizes.}
    \label{figure:soc:araos:fmatmul}
    \Description[\gls{OS} performance overhead for a matrix multiplication on different problem sizes.]{\gls{OS} performance overhead for a matrix multiplication on different problem sizes.}
\end{figure*}
As shown in \Cref{figure:soc:araos:fmatmul}.[b,c,d], we divide the overhead contribution into three parts: 1) caused by CVA6 \gls{MMU} requests, 2) by Ara2's \gls{MMU} request, and a remaining part caused by \gls{MMU} time multiplexing overhead, page-table-walk cache pollution, etc. 
Most notably, \gls{MMU} contention conflicts and \gls{DTLB} misses do not always impact performance. Since vector instructions run for multiple cycles, Ara2's \gls{FPU} computation can overlap and hide the stalls from \gls{DTLB} misses. Ara2 hides most of the stalls in our experiment, resulting in less than 3.5\% performance overhead on all problem sizes for a \gls{DTLB} with at least 16 entries.

A larger \gls{DTLB} yields better performance and also reduces \gls{OS} interference since the whole dataset of the application can fit in the pages addressed by the cached \glspl{PTE}. 
As we approach 128 \glspl{PTE}, all the pages that host the workloads can be cached in the \gls{DTLB}, virtually eliminating the overhead due to \gls{TLB} misses. However, due to the non-optimal pseudo-least-recently-used replacement policy of the \gls{DTLB}, some misses still occur, with an overhead below 1\%. The three matrix multiplication datasets can be contained in 6, 24, and 96 4-KiB pages. When running the same application on different problem sizes, larger problems need more \gls{DTLB} entries to reach their performance peak by eliminating the \gls{DTLB}-miss overhead (16, 32, 128 \glspl{PTE}, respectively). We can also observe that the \gls{DTLB} CVA6 overhead decreases when the program size increases, as longer vectors hide CVA6 stalls when loading scalar data elements.

\subsection{RiVEC benchmark suite}
\label{subs:rivec}
In this Section, we evaluate AraOS' performance using the RiVEC benchmark suite for \gls{RVV} processors\footnote{\url{https://github.com/RALC88/riscv-vectorized-benchmark-suite}, commit: 32a14dd}. 
Thus far, its adoption has been limited, probably because of its use of \gls{OS}-related functions, such as \texttt{malloc}. With \gls{OS} support, we can run RiVEC on AraOS.

\begin{table*}[ht]
\footnotesize
\caption{RiVEC benchmark suite performance on AraOS at 50 MHz. S: scalar (runtime in seconds). V: vector (performance multiplier w.r.t. scalar runtime). Vu: vector, with unordered reductions instead of ordered ones. Red: speedup lower than 1$\times$.}
\begin{tabular}{lrrrrrrrrrrrr}
\hline
\multicolumn{1}{c}{} & \multicolumn{3}{c}{\textbf{simtiny}} & \multicolumn{3}{c}{\textbf{simsmall}} & \multicolumn{3}{c}{\textbf{simmedium}} & \multicolumn{3}{c}{\textbf{simlarge}} \\ \cline{2-13} 
\multicolumn{1}{c}{\multirow{-2}{*}{Perf. 50MHz}} & \multicolumn{1}{c}{\textbf{S (s)}} & \multicolumn{1}{c}{\textbf{V ($\times$)}} & \multicolumn{1}{c}{\textbf{Vu ($\times$)}} & \multicolumn{1}{c}{\textbf{S (s)}} & \multicolumn{1}{c}{\textbf{V ($\times$)}} & \multicolumn{1}{c}{\textbf{Vu ($\times$)}} & \multicolumn{1}{c}{\textbf{S (s)}} & \multicolumn{1}{c}{\textbf{V ($\times$)}} & \multicolumn{1}{c}{\textbf{Vu ($\times$)}} & \multicolumn{1}{c}{\textbf{S (s)}} & \multicolumn{1}{c}{\textbf{V ($\times$)}} & \multicolumn{1}{c}{\textbf{Vu ($\times$)}} \\ \hline
\rowcolor[HTML]{FFFFFF} 
\textbf{axpy} & 1.94E-01 & 4.34 & 4.34 & 3.86E-01 & 4.28 & 4.28 & 7.73E-01 & 4.27 & 4.27 & 1.54E+00 & 4.26 & 4.26 \\ \hline
\rowcolor[HTML]{FFFFFF} 
\textbf{blackscholes*} & 8.74E-01 & 8.38 & 8.38 & 6.99E+00 & 8.37 & 8.37 & 2.80E+01 & 8.39 & 8.39 & 1.17E+02 & 8.60 & 8.60 \\ \hline
\rowcolor[HTML]{FFFFFF} 
\textbf{canneal*} & 1.96E-02 & \cellcolor[HTML]{FFCCC9}0.49 & \cellcolor[HTML]{FFCCC9}0.52 & 1.75E+01 & \cellcolor[HTML]{FFCCC9}0.71 & \cellcolor[HTML]{FFCCC9}0.78 & 5.74E+01 & \cellcolor[HTML]{FFCCC9}0.71 & \cellcolor[HTML]{FFCCC9}0.78 & 1.26E+02 & \cellcolor[HTML]{FFCCC9}0.70 & \cellcolor[HTML]{FFCCC9}0.79 \\ \hline
\rowcolor[HTML]{FFFFFF} 
\textbf{jacobi-2d} & 3.97E-03 & 1.87 & 1.87 & 2.92E-01 & 3.94 & 3.94 & 1.21E+01 & 3.88 & 3.88 & 2.43E+02 & 3.88 & 3.88 \\ \hline
\rowcolor[HTML]{FFFFFF} 
\textbf{lavaMD} & 3.45E-02 & 2.96 & 4.56 & 2.13E+01 & 1.87 & 2.92 & 8.96E+01 & 1.91 & 2.99 & 4.80E+02 & 1.91 & 2.99 \\ \hline
\rowcolor[HTML]{FFFFFF} 
\textbf{matmul} & 2.24E-01 & 2.72 & 3.76 & 1.82E+00 & 3.02 & 3.56 & 1.55E+01 & 3.20 & 3.47 & 2.02E+02 & 3.29 & 3.37 \\ \hline
\rowcolor[HTML]{FFFFFF} 
\textbf{particlefilter*} & 9.97E-02 & 1.08 & 1.08 & 2.62E+00 & 1.14 & 1.14 & 5.60E+01 & 1.65 & 1.65 & 3.33E+02 & 2.00 & 2.00 \\ \hline
\rowcolor[HTML]{FFFFFF} 
\textbf{pathfinder} & 7.46E-02 & 3.72 & 3.72 & 1.06E+01 & 7.07 & 7.07 & 4.40E+01 & 6.61 & 6.61 & 1.77E+02 & 6.51 & 6.51 \\ \hline
\rowcolor[HTML]{FFFFFF} 
\textbf{somier} & 1.36E+01 & 3.35 & 3.35 & 4.10E+01 & 3.36 & 3.36 & 1.17E+02 & 3.47 & 3.47 & 3.49E+02 & 3.44 & 3.44 \\ \hline
\rowcolor[HTML]{FFFFFF} 
\textbf{spmv**} & 4.23E-04 & \cellcolor[HTML]{FFCCC9}0.95 & 1.04 & 1.34E-01 & 1.42 & 1.89 & 0.14E+00 & 1.80 & 2.23 & 0.14E+00 & 1.80 & 2.23 \\ \hline 
\cellcolor[HTML]{FFFFFF}\textbf{streamcluster} & \cellcolor[HTML]{FFFFFF}4.08E+00 & \cellcolor[HTML]{FFFFFF}1.97 & \cellcolor[HTML]{FFFFFF}4.18 & 3.30E+02 & 1.94 & 3.72 & 6.10E+02 & 1.93 & 3.59 & 6.10E+02 & 1.93 & 3.59 \\ \hline
\rowcolor[HTML]{FFFFFF} 
\textbf{swaptions} & 1.84E+00 & 2.60 & 2.61 & 1.41E+01 & 2.68 & 2.67 & 1.12E+02 & 2.67 & 2.67 & 4.49E+02 & 2.66 & 2.65 \\ \hline
     \multicolumn{13}{l}{* Numerical mismatch in results. ** Med/large datasets too large for storage. \texttt{Na5} matrix used instead of \texttt{venkat25} and \texttt{poisson3Db}.}\\
\end{tabular}
\label{tab:rivec}
\end{table*}

\Cref{tab:rivec} shows the speedup of the vectorized code against the scalar code. We also add performance results by replacing the ordered reduction sums (verification) with unordered ones (benchmarking).
We expect the 2-lane AraOS to speed up the scalar code by at least $2\times$. This happens on 11/12 applications using the \texttt{simlarge} problem size, except for \texttt{canneal}.
These speedups are conservative since AraOS can process smaller or poorly vectorizable applications with the scalar core to avoid slowdowns.
Under this hypothesis, the average (geomean) speedups for increasing problem sizes are $2.7\times$, 3.0$\times$, 3.2$\times$, and 3.2$\times$.
With a 2.3$\times$ higher LUT utilization, this translates into system-level area efficiency gains from +17\% to +39\%.

The vectorized \texttt{canneal} is significantly slower than its scalar version on AraOS for multiple reasons. First, as reported in \cite{10.1145/3422667}, the exposed vectors are short (from 5 to 22 elements, 10 on average)—a condition in which Ara2 is inefficient. More importantly, \texttt{canneal} reinterprets the content of a vector register with two different \glspl{EW}. This condition triggers a reshuffle in Ara2 \cite{ara2perotti24} at every iteration to adapt the byte layout to the new \gls{EW}. This is especially detrimental when operating on short vectors since the reshuffle is performed on the entire vector register length, even if the current vector length is way shorter. Also, Ara2's conservative dependency checks prevent the following load instruction from chaining the reshuffle, whose latency cannot be amortized.
Also, \texttt{canneal} and \texttt{spmv} underperform compared to the scalar-only benchmark because of their reliance on indexed memory operations that are not optimized on AraOS, which pays the latency of a dedicated address translation on each vector element to ensure precise exceptions.
In \texttt{spmv}, unordered reductions and more \glspl{NZE} per \gls{NER} (\ie longer vectors) boost performance. As the problem size grows, the \glspl{NZE} per \gls{NER} are $\sim$5 (tiny), $\sim$21 (small), and $\sim$27 (med, large), and the speedup increases from 0.95$\times$ (min) to 2.23$\times$ (max).

\section{Conclusions}
In this paper, we add \gls{OS} support to the Ara2 vector processor (AraOS), integrate it in an open-source \gls{SoC}, and prototype it on \gls{FPGA}.
We characterize the effect of an \gls{OS} on vector performance by analyzing the impact of scheduler interrupts, context switches, and \gls{DTLB} size on key matrix multiplication kernels.
Vector execution helps hide part of the stalls caused by \gls{DTLB} misses and \gls{MMU} contention, keeping the virtual memory performance overhead over the bare-metal baseline below 3.5\% starting from 16 \gls{DTLB} entries. 
Finally, \gls{OS} support allows us to benchmark AraOS with the open-source RiVEC benchmark suite, showing an average 3.2$\times$ speedup and up to +39\% area efficiency over the scalar baseline. 

\begin{acks}
This work was supported in part through the ISOLDE (101112274) project that received funding from the HORIZON CHIPS-JU programme, in part
by the PON “Ricerca e Innovazione” 2014-2020, Azione IV.5, DM n.1061, Italian Ministry of University and Research,
and received funding from the Swiss State Secretariat for Education, Research, and Innovation (SERI) under the SwissChips initiative.
\end{acks}

\bibliographystyle{ACM-Reference-Format}
\bibliography{main}


\end{document}